# Re-entrant charge order in overdoped $(Bi,Pb)_{2.12}Sr_{1.88}CuO_{6+\delta}$ outside the pseudogap regime

Y. Y. Peng[1], R. Fumagalli[1], Y. Ding[2], M. Minola[3], S. Caprara[4,5], D. Betto[6], G. M. De Luca[7,8], K. Kummer[6], E. Lefrançois[3], M. Salluzzo[8], H. Suzuki[3], M. Le Tacon[9], X. J. Zhou[2], N. B. Brookes[6], B. Keimer[3], L. Braicovich[1,10], M. Grilli[4,5] and G. Ghiringhelli[1,10]*

[1] Dipartimento di Fisica, Politecnico di Milano, Piazza Leonardo da Vinci 32, I-20133 Milano, Italy. [2] Beijing National Laboratory for Condensed Matter Physics, Institute of Physics, Chinese Academy of Sciences, Beijing 100190, China. [3] Max-Planck-Institut für Festkörperforschung, Heisenbergstraße 1, D-70569 Stuttgart, Germany. [4] Dipartimento di Fisica, Università di Roma "La Sapienza", P.le Aldo Moro 5, 00185 Roma, Italy [5] CNR-ISC, via dei Taurini, 19 - 00185 Roma, Italy. [6] ESRF, The European Synchrotron, 71 Avenue des Martyrs, Grenoble, France. [7] Dipartimento di Fisica "E. Pancini", Università di Napoli Federico II, Complesso Monte Sant'Angelo, Via Cinthia 80126 Napoli, Italy. [8] CNR-SPIN, Complesso MonteSantangelo - Via Cinthia, I-80126 Napoli, Italy. [9] Institute of Solid State Physics (IFP), Karlsruhe Institute of Technology, D-76021 Karlsruhe, Germany. [10] CNR-SPIN, Dipartimento di Fisica, Politecnico di Milano, Piazza Leonardo da Vinci 32, I-20133 Milano, Italy.

* Correspondence to: giacomo.ghiringhelli@polimi.it

**Charge modulations are considered as a leading competitor of high-temperature superconductivity in the underdoped cuprates, and their relationship to Fermi surface reconstructions and to the pseudogap state is an important subject of current research. Overdoped cuprates, on the other hand, are widely regarded as conventional Fermi liquids without collective electronic order. For the overdoped $(Bi,Pb)_{2.12}Sr_{1.88}CuO_{6+\delta}$ (Bi2201) high-temperature superconductor, here we report resonant x-ray scattering measurements revealing incommensurate charge order reflections, with correlation lengths of 40-60 lattice units, that persist up to at least 250K. Charge order is markedly more robust in the overdoped than underdoped regime but the incommensurate wave vectors follow a common trend; moreover it coexists with a single, unreconstructed Fermi surface, without pseudogap or nesting features, as determined from angle-resolved photoemission spectroscopy. This re-entrant charge order is reproduced by model calculations that consider a strong van Hove singularity within a Fermi liquid framework.**

High-temperature superconductivity emerges upon doping of holes or electrons into Mott-insulating copper oxides. The strong electronic correlations responsible for Mott localization in the parent compounds generate various competing instabilities in the underdoped regime[1]. As recent experiments have established charge order (CO) as a universal feature of moderately doped cuprates, its relationship to the ubiquitous "pseudogap" phenomenon has been at the focus of several studies. Early evidence of charge order had come from La-based cuprates, where charge "stripes" were observed near the doping level $p$=1/8 holes per Cu (refs. [2,3,4]). Recent resonant x-ray scattering (RXS) experiments also revealed incommensurate charge order competing with superconductivity in $YBa_2Cu_3O_{6+x}$ (Y123) and in Hg- and Bi-based cuprates[5,6,7,8,9,10,11,12]. A detailed comparison of the x-ray data to angle-resolved photoemission spectroscopy (ARPES) data on Bi-based cuprates showed that the onset temperature of CO is close to the pseudogap temperature $T^*$, and that its wave-vector is comparable to the distance between the Fermi arc tips, therefore hinting at a link between CO and the pseudogap in hole-doped systems[13,14]. Also in Y123 the CO onset temperature appears to be always lower than $T^*$, whereas in electron-doped cuprates CO extends well above $T^*$, with an onset temperature close to that of the antiferromagnetic fluctuations, thus suggesting a possible connection between the two[15].

Based on these observations, it is now widely assumed that CO is restricted to the underdoped region of the phase diagram characterized by the pseudogap. However, various instabilities have also been predicted near the "Lifshitz point" in the overdoped regime, where a van-Hove singularity (vHs) in the electronic density of states moves across the Fermi level and the geometry of the Fermi surface changes from hole-like to electron-like. The Bi2201 system is well suited to test these predictions, because the doping level can be tuned over a wide range, well into the overdoped regime and, thanks to very distant $CuO_2$ layers, it generates a single Fermi surface. The single Lifshitz point resulting from this electronic structure greatly facilitates the quantitative correlation between data generated by ARPES and RXS. Additionally, the vHs is particularly strong due to the pronounced 2D character of this system[16], where the superconducting planes are highly decoupled.

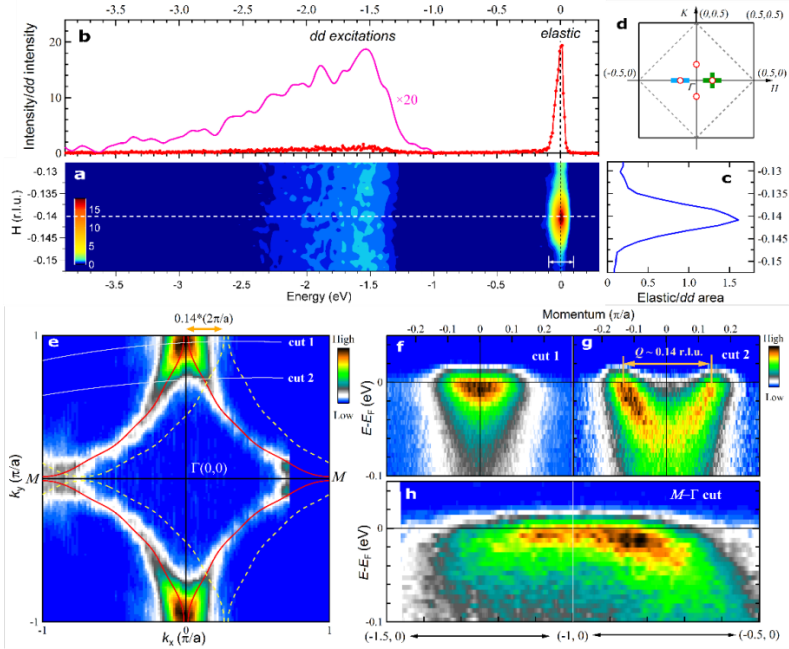

**Figure 1 Observation of a quasi-elastic peak by RIXS in overdoped $(Bi,Pb)_{2.12}Sr_{1.88}CuO_{6+\delta}$ ($T_c$=11K, $p\sim0.215$).** a, Energy/momentum intensity map of RIXS spectra along the (-0.5,0) - (0,0) symmetry direction, indicated by the thick blue line in panel (d). The data were taken with σ-polarized incident light at 20K. The RIXS spectrum at $H \simeq$ -0.14 r.l.u. indicated by the white dashed line is shown in (b). The *dd* excitations are shown additionally after smoothing and multiplied by 20 to be comparable to the elastic peak. c, The quasielastic RIXS intensity is given by the integral around $E \sim 0$, as indicated in (a). d, Reciprocal-space image. The hollow red circles indicate the observed quasi-elastic peak. e, Photoemission intensity at the Fermi energy ($E_F$) as a function of momenta $k_x$ and $k_y$ for OD11K at 20K. It is obtained by integrating within a (-10 meV, 10 meV) energy window and symmetrizing the original data with respect to the (-π/a,0) - (π/a,0) line. The red lines, obtained by tight-binding fitting to the data, serve as a guide to the eyes. The dashed yellow lines indicate the Fermi surface shifted horizontally by $Q_{CO}\simeq 0.14$ r.l.u.. f, g, Electronic dispersions for the cuts (indicated by the white lines in e). h, Electronic dispersion for the cut along $M$-Γ direction near the Brillouin zone boundary.

Here we present resonant *inelastic* x-ray scattering (RIXS) data that display sharp, intense incommensurate CO diffraction peaks in overdoped Bi2201 over a range of doping levels that spans the Lifshitz point and the end point of the superconducting dome. The smooth decrease of the CO vector with doping demonstrates the ubiquity of charge order across the entire phase diagram. Together with ARPES data on the same samples, our results challenge models that posit an essential link between CO and the pseudogap.

We performed RIXS measurements at the Cu $L_3$ edge on overdoped Bi2201 at four different doping levels (see Methods). We will hereafter use the common notations for the in-plane wave vector $\mathbf{Q}_{//}$, the pseudo-tetragonal reciprocal lattice units (r.l.u.) $2\pi/a=2\pi/b=1$ (with $a \simeq b \simeq 3.83$ Å), and the reciprocal space indices ($H,K,L$). Figure 1a shows the energy/momentum intensity maps for OD11K ($T_c$=11K, $p\sim0.215$) along the $H$ direction. The inelastic features in the [-3,-1] eV range, due to inter-orbital transitions (*dd* excitations)[17], depend weakly on momentum, whereas the response centred at zero energy loss exhibits a pronounced maximum at $\mathbf{Q}_{//}= (-0.14, 0)$, shown in Fig. 1c. The RIXS spectrum at $H \simeq -0.14$ r.l.u. (Fig. 1b) is dominated by the elastic peak, which is ~20 times more intense than the *dd* excitations. We found the peak to be elastic within our experimental uncertainty (∼10 meV), and will refer to this feature as a "resonant *elastic* x-ray scattering" (REXS) peak.

Figure 1e shows the Fermi surface (FS) of OD11K measured by ARPES at 20K. We found no replicas of the large Fermi surface, and its shape offers no parallel segments suitable for a good nesting at the CO wave vector (0.14,0): the shifted FS exhibits a point-like crossing with the original one. In the antinodal region the band lies very close to the Fermi level energy ($E_F$), with no space for nesting (Fig. 1f). A cut through the FS points separated by $\mathbf{Q}_{//}$ (Fig. 1g) shows no gap opening at $E_F$, which one would expect in a folded FS due to charge ordering[18]. On the other hand, a strong van Hove singularity is located slightly below $E_F$ at the M point in the Brillouin zone boundary (Fig. 1h and Supplementary Fig. S5).

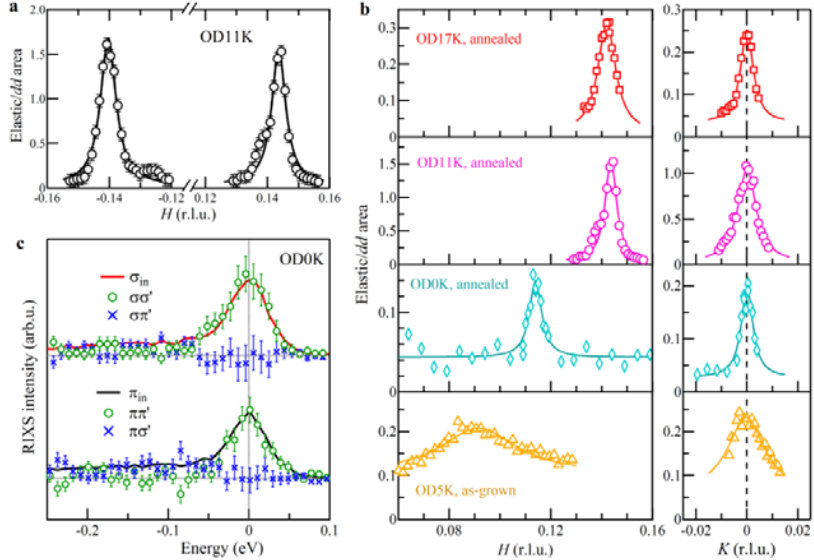

**Figure 2 Doping and polarization dependence of the REXS peak in (Bi,Pb)$_{2.12}$Sr$_{1.88}$CuO$_{6+\delta}$.** a, REXS intensity for positive and negative $H$, using σ polarization. b, $H$ and $K$ cuts at 20K as indicated by the green cross in Fig. 1d, at 4 doping levels. Solid lines are Lorentzian peak fits to the data with a constant background. c, Polarization resolved measurements for OD0K ($p\simeq0.23$) taken at $H \simeq 0.115$ r.l.u. with incident π- and σ-polarized light. Statistical error bars are calculated from the number of counts (see Supplementary Information).

The REXS peak was found at both positive and negative $H$ with similar intensity, as shown in Fig. 2a, and along both (1,0) and (0,1) directions. On the contrary, we could not detect it along the (1,1) direction (Supplementary Fig. S7), in agreement with prior work on CO in underdoped cuprates[19,20]. Figure 2b provides the doping dependence of the REXS peak, with cuts along both $H$ and $K$ directions, as indicated by the green cross in Fig. 1d. It is important to note that three doping levels (OD17K, $p\sim0.205$; OD11K, $p\sim0.215$; OD0K, $T_c<2$ K, $p\sim0.23$) were obtained by post-annealing treatments of the as-grown OD5K ($p\sim0.225$). Therefore the latter is expected to have a higher degree of structural disorder, as confirmed by the much larger and anisotropic width of the REXS peak and by the broader superconducting transition (Supplementary Fig. S1); moreover, the $Q_{CO}$ of OD5K falls out of the smooth trend set by the three other samples.

To determine whether the REXS peak arises from spin or charge order, we have exploited the polarimeter of our RIXS facility[21]. Spin-related scattering implies a 90° rotation of the photon polarization[22], whereas pure charge scattering, without spin-flip, conserves the photon polarization. As shown in Fig. 2c, the REXS peak is purely polarization-conserving: the σσ' and ππ' channels comprise the entire signal, whereas the πσ' and σπ' intensities are zero within the experimental error bars.

To clarify whether the peak originates from lattice displacements or from a modulation of the valence electron charge density, we checked its resonant behaviour. A modulation of the lattice would generate sizable Bragg reflections also under non-resonant conditions, whereas pure charge order is detectable only at resonance. Figure 3a shows the incident photon energy dependence of the REXS peak, which closely resembles the spectra of the Cu $L_3$ x-ray absorption spectrum (XAS). Unlike YBCO, which has two inequivalent Cu sites contributing to the XAS spectra[6], in Bi2201 the Cu $L_3$ absorption peak is fully due to the Cu in the $CuO_2$ planes, and the strictly resonant peak can thus be unequivocally assigned to modulations of the charge density in the $CuO_2$ planes. Figure 3b shows the REXS scans measured at the maximum of the Cu $L_3$ absorption profile and at three photon energies above the maximum. While the REXS peak intensity decreases upon detuning from the resonance, its position and width are unchanged. Within the error, this photon energy dependence is identical to the resonance of CO in underdoped Bi2201. The only difference is that the signal here is still visible 2 eV above the Cu $L_3$ edge, where it was no longer discernible in the earlier resonant scattering experiments[13]. This is due to the higher signal-to-background ratio in our experiments, which were carried out with energy analysis of the scattered beam.

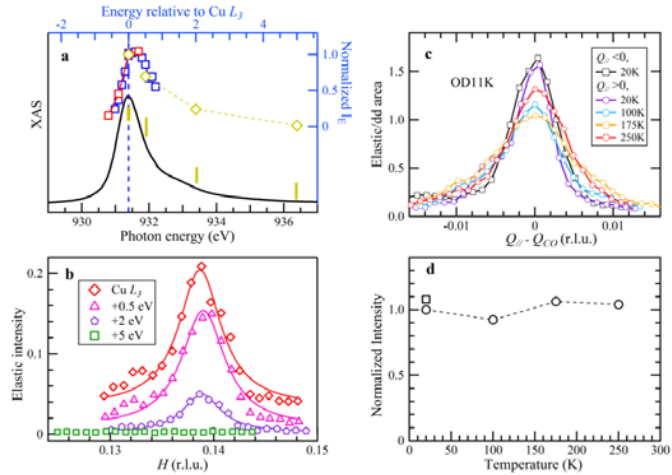

**Figure 3 Energy and temperature dependence of the REXS peak in $(Bi,Pb)_{2.12}Sr_{1.88}CuO_{6+\delta}$.** a, (Left/Bottom) XAS spectra of UD17K with σ polarization at normal incidence. (Right/Top) Incident energy dependence of the REXS intensity, normalized to the value at XAS peak: hollow blue (red) squares for π (σ) polarization of OD11K, brown diamonds for σ polarization of OD17K. b, REXS scans along $H$ direction for OD17K at 20K, at selected incident energies indicated by the brown lines in (a). Solid lines are Lorentzian peak fits to the data with a constant background. c, Comparison of REXS scans at selected temperatures for OD11K with $Q_{CO} \simeq \pm 0.14$ ($\pm 0.005$) r.l.u.. d, $T$-dependence of the charge order intensity, normalised to the value at 20K at $Q_{//}>0$.

Figure 3c shows the temperature dependence of the REXS peak in OD11K. Although the peak broadens slightly as the temperature is raised, its integrated intensity is almost temperature independent up to 250K (Fig. 3d), indicating that the onset of charge order occurs well above 250K. Since $T^*$ of OD11K is approximately zero[23,24], as confirmed by the absence of gap at the antinode at 20K (Fig. 1f), this means that the CO peak is present in the absence of the pseudogap. Additional measurements on OD17K confirmed that the REXS peak is present at 20K and at 100K (Supplementary Fig. S8), i.e. both below and above $T^*\sim 60K$ (ref. 25). Due to the relatively low $T_c$ of our samples, all falling below the base temperature reachable by the experimental apparatus (~20K), we could not look at the CO peak in the superconducting state to determine the interplay between charge order and superconductivity. However, when approaching $T_c$ from above, we did not observe changes in the CO peak that would indicate the proximity of superconductivity.

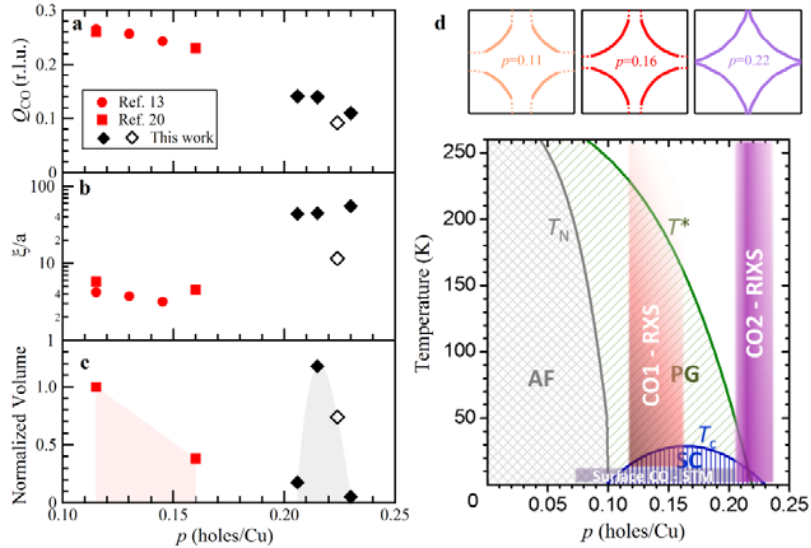

**Figure 4 Doping dependence of the charge order signal in $(Bi,Pb)_{2.12}Sr_{1.88}CuO_{6+\delta}$ and the corresponding phase diagram.** a-c, Doping dependence of the CO wave vector, correlation length and volume. Data from RXS[13] are included; black diamonds for annealed (solid) and as grown (hollow) samples. Error bars are smaller than symbol dimensions. d, The phase diagram of the charge order in Bi2201: it shows the antiferromagnetic region (AF) defined by $T_N$, superconducting region (SC) defined by $T_c$, and the pseudogap region (PG) defined by $T^*$, which are reproduced from NMR measurements[24]. The grey shaded area for the checkerboard charge order observed with STM from $p=0.07$ (ref. 26) to $p=0.21$ (ref. 27); the red shaded area for the CO measured with x-ray[13,20]; the violet shaded area denotes the region of the newly discovered CO. The Fermi surfaces at selected doping $p=0.11$, 0.16 and 0.22 are schematically depicted on the top panels, breaking into Fermi arcs at $T<T^*$.

Figures 4a to c present an overview of the wave vector of the REXS peak, $Q_{CO}$, as well as its correlation length and its integrated intensity ("volume") as functions of doping, including earlier data on underdoped Bi2201 (refs. 13,20). The newly discovered $Q_{CO}$ is approximately half of that of the underdoped samples, smoothly extending the known negative slope of $Q_{CO}(p)$ to the overdoped region, hinting at strong commonality of CO in the underdoped and overdoped regimes. As already mentioned, the peak intensity (normalized to the *dd* excitations) is much higher in the overdoped samples. This is combined with a much smaller width in Q-space: the charge order is rather long-ranged in the overdoped region, with the only exception of the as-grown OD5K whose REXS peak is relatively weak and broad. In the three other overdoped samples the correlation lengths $\xi_{H,K}$ are in the range of $40a$ to $60a$, an order of magnitude larger than those of underdoped Bi2201 (~$6a$). It is noteworthy that this long-range correlation of charge order is comparable with the stripe-order in $La_{1.875}Ba_{0.125}CuO_4$ (refs. 2,3,4) and the field-induced CO in YBCO (refs. 9,10). The latter similarity hints at a competition between CO and SC at high doping too.

For a more quantitative assessment we determined the k-space volume of the REXS peak, which is proportional to $\xi_H^{-1} \times \xi_K^{-1} \times d$, where $d$ is the height of the REXS peak after subtracting the background from the curves of Fig. 2b. For simplicity, we present in Fig. 4c the volume normalized to the value at $p \simeq 0.115$. We notice that the CO integral intensity has two maxima, one around $p \sim 0.115$ (but no RXS data are available for lower doping) and one around $p \sim 0.215$. Interestingly the peak volumes of these two maxima are quite comparable. Moreover in the as-grown, more disordered sample the volume of the CO peak is comparable to the other overdoped samples, indicating that the charge instability is robust versus structural disorder, and confirming that what we have observed is not a trivial lattice superstructure.

Figure 4d shows the extended phase diagram of charge order in Bi2201, including also the checkerboard-like charge order that was observed by scanning tunnelling microscopy (STM) from the insulating state ($p \simeq 0.07$)[26] up to OD15K ($p \simeq 0.21$)[27], and characterized by a doping-independent $Q_{CO} \sim 0.25$ r.l.u.. On the other hand, RXS measurements revealed a short-ranged CO in the pseudogap state up to optimal doping, and a long-ranged charge order, with small $T$-dependence between 20 K and 250K, outside the pseudogap region up to $p \simeq 0.23$. Given the different charge order wave vectors determined from STM and RXS, their relations remain to be further studied. The schematic Fermi surfaces of Bi2201 at three selected dopings ($p \simeq 0.11$, 0.16 and 0.22) are shown on the top panels. A hole-like Fermi surface centred at ($\pm\pi/a, \pm\pi/a$) is observed in a broad doping range; it grows in size with doping and breaks into Fermi arcs below $T^*$ (refs. 28,29). It eventually transforms into an electron-like FS centred at (0,0) in the case of OD0K, where the Lifshitz transition takes place in Bi2201 (ref. 26).

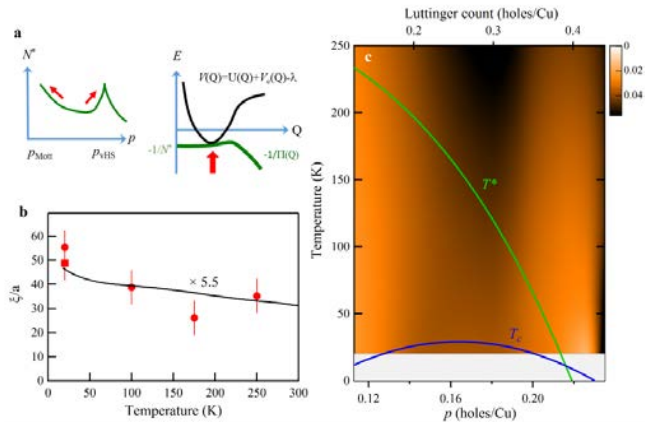

**Figure 5 Theoretical (strongly-correlated) RPA results.** a, When the quasiparticle DOS ($N^*$) grows, the whole $-1/\Pi(Q)$ curve (thick green line) increases and becomes tangent to the effective interaction curve $V(Q)$ (black line), therefore fulfilling the instability condition. This can happen either at low doping or at high doping. This latter DOS growth is due to the vHs that promotes the charge instability generating the re-entrant CO behaviour. b, Correlation length for OD11K (symbols) and theory (black line). The theory curve is multiplied by 5.5 to be comparable to experimental data. c, Phase diagram reporting the characteristic energy (mass) $m \sim \square^{-2}$ of the CO fluctuations above 20K. A smaller mass indicates a longer-ranged CO correlation (but long range order is never reached as long as $m>0$). Luttinger counts are obtained from the Fermi surface measured by ARPES and used in the calculations (see Methods).

The discovery of CO in the overdoped superconducting region establishes a unified phenomenology on both sides of the superconducting dome, and poses intriguing questions regarding the mechanism underlying this phenomenon. Spin instabilities have only been reported close to the Mott-insulating state and are therefore unlikely to play a major role in the mechanism driving the CO in the overdoped regime. Various instabilities have been predicted by other approaches for doping levels close to the vHs where a large density of states moves across the Fermi level[30,31,32,33,34]. However, these instabilities are typically associated with nesting features of the Fermi surfaces, which are not observed in our experiments.

Figure 5 shows the basic ingredients and results of an alternative, Fermi-liquid based theoretical approach that captures several key properties of the re-entrant CO. The strongly correlated character of cuprates is encoded in the mass renormalization of the Fermi-liquid quasiparticles and in a momentum-dependent residual interaction $V(\mathbf{Q}_\parallel)$ (see the Methods section for details). Within a random phase approximation (RPA), a charge instability occurs when $1+V(\mathbf{Q}_\parallel)\Pi(\mathbf{Q}_\parallel)$, the denominator of the charge susceptibility, vanishes, i.e. for those $\mathbf{Q}_{co}$ where $V(\mathbf{Q}_{co})$ equals $-1/\Pi(\mathbf{Q}_{co})$, with $\Pi(\mathbf{Q}_\parallel)$ being the Lindhard polarization function of the quasiparticles (Fig.5a). This gives rise to a CO quantum critical point (QCP) around optimal doping, which is supported by prior experimental and theoretical work [35,35,36,35]. However, the nearly 2D character of cuprates favours strong fluctuations that tend to destroy long-range charge order[34,36,37]; the QCP is thus shifted to smaller doping, and at nonzero temperatures the long range order is replaced by nearly critical CO fluctuations. The phase diagram of Fig. 5c is constructed from our ARPES results on Bi2201 (see Methods); it reports the characteristic energy of the (overdamped) CO fluctuations, expressed as their mass $m$ (with the correlation length $\xi \sim m^{-1/2}$). This energy is nonzero over the whole doping and temperature range but, for Bi2201, it approaches zero both in the underdoped and overdoped regimes. In the RPA approach, the re-entrant CO can thus be explained by the Lindhard function having the quasiparticle density of states $N^* = \Pi(\mathbf{0})$ as a characteristic scale. When $N^*$ increases due to the vHs (Fig. 5a), $-1/\Pi(\mathbf{Q}_\parallel)$ increases and the instability condition is met. This approach also reproduces the temperature-dependence of the correlation length (Fig. 5b), although the agreement requires a prefactor that could be further adjusted in the theory.

Our model explains not only the CO around the vHs but, within the same scheme, it accounts for the generic presence of CO in the underdoped region as well[35,36,37]. Moreover, the fact that the CO wave vector is not tied to any nesting features of the Fermi surface is compatible with our Fermi liquid model, where it is mostly due to the momentum dependence of the residual interaction, whereas this observation is difficult to reconcile with alternative schemes[34].

Future work will have to address the detailed structural and electronic texture: a re-entrant CO due to the vHs might be less universal than CO in the underdoped regime, because vHs strongly varies among cuprate families, being magnified in highly anisotropic compounds, such as single layer Bi- and Tl-based cuprates. In any case, our discovery of a strong electronic instability requires a reassessment of the commonalities and differences of cuprates across the optimal doping level and the quantum critical point nearby.

# Methods

**Sample characterisation**

Single crystals of $(Bi,Pb)_{2.12}Sr_{1.88}CuO_{6+\delta}$ are grown by the travelling solvent floating zone method. The sample growth and characterization methods have been reported previously[38]. The as-grown single crystals with nominal composition $Pb_{0.25}Bi_{1.87}Sr_{1.88}CuO_{6+\delta}$ were post-annealed in different atmospheres including vacuum and high pressured oxygen at different temperatures (500$^0$C ~ 600$^0$C) in order to adjust the doping level to change $T_c$ and to make the samples homogeneous. Supplementary Figure S1 shows the temperature dependence of magnetization for as-grown Bi2201 single crystals and for samples after various annealing processes. The as-grown OD5K showed a broad transition width of more than 3K. After annealing, the magnetization of OD17K and OD11K showed a sharp transition width ~1K. The OD0K showed no superconducting transition down to 2K, which was the minimum temperature of our apparatus, thus we indexed it as OD0K.

There are several methods to determine the doping $p$ in Bi2201, which was discussed in detail in the supplementary materials of ref. 27. Ando et al.[39] proposed an empirical relation between $T_c$ and the hole concentration $p$ in Bi2201 based on a comparison of Hall measurements in several cuprates:

$$T_c = T_{c,max}[1-278(p-0.16)^2]$$

which we have used to determine $p$ for our four samples from the measured $T_c$ in the main text.

The doping $p$ can be also determined as Luttinger count from Fermi surface area according to Luttinger's theorem[40]. The trend of Luttinger count vs. $p$ determined from $T_c$ is shown in Supplementary Fig. S6. To be noted, our theory calculations have used the doping $p$ of Luttinger count, which corresponds to the real band structures measured by ARPES.

**ARPES measurements**

The ARPES measurements were carried out on the angle-resolved photoemission system with a Helium discharge lamp in the Institute of Physics, Chinese Academy of Sciences, China. The photon energy was 21.218 eV and the energy resolution was set to 10 meV, and the angular resolution was 0.3 degree. The samples were cleaved *in situ* and measured under ultrahigh vacuum, pressure lower than $6\times10^{-11}$ mbar. The Fermi level is referenced by measuring on the Fermi edge of a clean polycrystalline gold that is electrically connected to the sample.

**RIXS measurements**

The RIXS measurements were performed at the ID32 of ESRF (The European Synchrotron, France) using the new high-resolution ERIXS spectrometer. The resonant conditions were achieved by tuning the energy of the incident x-ray to the maximum of the Cu $L_3$ absorption peak, around 931 eV. The total instrumental energy resolution was set at 65 meV, determined as FWHM of the non-resonant diffuse scattering from the silver paint. The samples were cleaved out-of-vacuum to expose a fresh surface. The XAS measurements were made at the ID32 of ESRF. The RIXS experimental geometry is shown in Supplementary Fig. S3. X-rays are incident on the sample surface and scattered by an angle 2θ. Reciprocal lattice units (r.l.u.) were defined by using the pseudo-tetragonal unit cell with $a = b = 3.83$ Å and $c = 24.54$ Å, where the axis c is normal to the cleaved sample surface. The sample can be rotated azimuthally around the $c$ axis to choose the in-plane wave vector component. Data in the text were taken with 2θ = 149°, giving $|\mathbf{Q}| = 0.85$ Å$^{-1}$, which allows one to cover the whole first Brillouin zone along the [100] direction (~0.82 Å$^{-1}$). Here, the negative (positive) $Q_\parallel$ corresponds to grazing-incidence (grazing-emission) geometry. Each RIXS spectrum was measured for 1 minute (sum of individual spectra of 5 seconds). The quasi-elastic intensity was determined by the integral of $0 \pm 0.1$ eV. That intensity was normalized to the integral of *dd* excitations in the RIXS spectrum between 1 and 3 eV energy loss. For the polarimeter measurements, the spectrum without polarimeter was measured for 15 min and the spectrum with polarimeter was measured for 50 min. The total instrumental energy resolution was set at ~100 meV.

**The theoretical model and calculations**

Our theoretical approach is based on a Fermi liquid scheme. The strongly correlated character of cuprates is encoded in the mass renormalization of the quasiparticles (QP) and an effective momentum-dependent residual interaction. Starting from an infinite-$U$ Hubbard model, a standard slave-boson approach[35] can be used to provide the doping dependence of the QP mass $m^* = m_{el}/p$, (where $p$ is the doping, so that is $m^* \sim 3\text{-}5 m_{el}$ in the region of interest) and the residual interaction

$$V(\mathbf{Q}) = U(\mathbf{Q}) - \lambda_\mathbf{Q} + V_c(\mathbf{Q}) \tag{1}$$

where $U(\mathbf{Q}) \simeq \frac{U_0 + U_1 q^2}{p}$ is the residual repulsion stemming from the original bare large Hubbard $U$, $\lambda_\mathbf{Q}$ is an attraction, which may arise from phonons or non-critical spin interaction (like, e.g., a nearest-neighbour magnetic coupling), and $V_c(\mathbf{Q})$ is the Coulombic repulsion (a more complete description of the approach can be found in the SM). In the Random Phase Approximation the charge response function can be written as

$$\chi(\mathbf{Q}) = \frac{\Pi(\mathbf{Q})}{1+V(\mathbf{Q})\Pi(\mathbf{Q})} \qquad (2)$$

where $\Pi(\mathbf{Q})$ is the Lindhard function of the QP and a charge instability at a given wave vector $\mathbf{Q}_{co}$ is obtained when its denominator vanishes $[1 + V(\mathbf{Q}_{co})\Pi(\mathbf{Q}_{co})] = 0$, that is $V(\mathbf{Q}_{co}) = -\frac{1}{\Pi(\mathbf{Q}_{co})}$. This condition is schematically represented in Fig. 5a (right panel). It must be noticed that *both direction and modulus of $\mathbf{Q}_{co}$ are not directly related to nesting features of the Fermi surface.*

The CO collective modes are obtained by expanding the denominator of $\chi(\mathbf{Q}, \omega)$ for small frequency and small deviations from the instability wave vector $\mathbf{Q}_{co}$, obtaining the CO propagator

$$D(\mathbf{Q},\omega) = \frac{g^2}{m + \nu(\mathbf{Q}) - i\omega - \frac{\omega^2}{\Omega}} \qquad (3)$$

$g$ is the coupling between the fermionic QP and the CO fluctuations; $\nu(\mathbf{Q}) \approx \nu |\mathbf{Q} - \mathbf{Q}_{co}|^2$ is the dispersion law of Landau-damped CO fluctuations ($\nu$ is an electronic energy scale since we work with dimensionless momenta, measured in inverse lattice spacing $1/a$). $\Omega$ is the energy scale above which the CO critical mode becomes less overdamped and more propagating[37]. The CO mass term $m = \nu\xi^{-2}$ is the most relevant parameter to determine criticality. According to standard theory of quantum critical phenomena[36,37,41], it should be corrected by the four-leg interaction $u$ between CO fluctuations so as to include the effect of CO fluctuations on the mean-field (RPA) CO instability. To summarize: the RPA analysis provides a mean-field-like mass $m_0(x,T)$, vanishing at a mean-field instability line $T^0(p)$, and below which the mean-field CO mass becomes negative. However, the fluctuations, whose effect is proportional to the (four-leg) CO-CO coupling $u$ may disrupt this mean-field long-range order and leave a state with CO fluctuations only (i.e. without CO long-range order). The (negative) mean-field mass $m_0$ is corrected by fluctuations by a term $\delta m$, whose complete analytical expression is given in the SM. Of course this effect is particularly pronounced in nearly 2D systems, where the CO transition (marked by a negative renormalized mass $m = m_0 + \delta m$) only occurs at finite temperatures when the inter-plane coupling $\nu_{3D}$ intervenes to introduce a weakly 3D character[36,37]. This finite temperature may even fall underneath the superconducting $T_c(p)$ dome, so that static order only becomes visible when superconductivity is weakened by, e.g. a magnetic field[9,10].

The quasiparticle band dispersion was given by

$$E(\mathbf{k}) = \varphi(p)\left[\frac{1}{2}t(\cos k_x + \cos k_y) + t'\cos k_x \cos k_y + \frac{1}{2}t''(\cos 2k_x + \cos 2k_y)\right.$$
$$\left. + \frac{1}{2}t'''(\cos k_x \cos 2k_y + \cos 2k_x \cos k_y)\right]$$

with *t= 0.828 eV, t'/t=–0.154, t''/t=0.164, t'''/t=–0.06, and φ(p)=p/0.375*. In this way the QP have band dispersion coinciding with the one determined by ARPES at the Lifshitz point, but acquire a narrower (broader) bandwidth when the doping decreases (increases).

As far as the interaction is concerned, the short range repulsion was taken as $U(\mathbf{Q}) \approx \frac{U_0 + U_1 Q^2}{p}$ with $U_0$=0.193 eV, $U_1$=0.083 eV, for doping $p \geq 0.315$, while the expression saturates at its $p$=0.315 value when $p$<0.315.
The attraction $\lambda_\mathbf{Q}$ was taken equal to 1.11 eV along the (0,1) direction of the Brillouin zone and gradually decreasing when moving towards the (1,1) direction, where it reached the value 0.91 eV. The strength of the screened Coulomb repulsion was taken as $V_c$=0.085 eV. In this way we find a RPA instability both at doping $p$<0.225 and at 0.325≤$p$≤0.425. Owing to the strongly 2D character of Bi2201, in the phase diagram of Fig. 5c, this long-range CO is spoiled by the mass correction due to CO fluctuations and never occurs for $T$>20K. Here we used $\nu = \frac{\Lambda}{\bar{Q}^2}$, with $\Lambda$=1.2 eV, $\bar{Q}$~0.05 r.l.u., $\Omega = 0.2$eV, and $u$=0.23 eV, yielding a finite positive mass $m$ (i.e. no true long-range order) over the whole phase diagram reported in Fig. 5c.

This work was supported by ERC-P-ReXS project (2016-0790) of the Fondazione CARIPLO, in Italy. M. M. was partially supported by the Alexander von Humboldt Foundation. XJZ thanks financial support from the National Natural Science Foundation of China (11334010 and 11534007), the National Key Research and Development Program of China (2016YFA0300300) and the Strategic Priority Research Program (B) of Chinese Academy of Sciences (XDB07020300). SC and MG acknowledge financial support from the Sapienza University Project No. C26A115HTN. The authors acknowledge insightful discussions with T. P. Devereaux, S. Kivelson, C. Di Castro, B. Moritz, and W. Metzner. The experimental data were collected at the beam line ID32 of the European Synchrotron (ESRF) in Grenoble (F) using the ERIXS spectrometer designed jointly by the ESRF and Politecnico di Milano.